%% file: main.tex
\documentclass{llncs}

\usepackage{xspace}
\usepackage{tikz}
\usetikzlibrary{shapes,arrows,positioning,calc}
\usetikzlibrary{shapes.misc}
\tikzset{cross/.style={cross out, draw=black, minimum size=2*(#1-\pgflinewidth), inner sep=0pt, outer sep=0pt},
cross/.default={5pt}}

\usepackage{tkz-euclide}
\usepackage{mathptmx}
\usepackage{proof}
\usepackage{amssymb}
\usepackage{url}

\newcommand{\vampire}{{\sc Vampire}}
\newcommand{\Vampire}{{\sc Vampire}}
\newcommand{\DC}{\ensuremath{\mathit{DC}}}
\newcommand{\CS}{\mathcal{C}}

\makeatletter
\renewcommand\paragraph{\@startsection{paragraph}{4}{\z@}%
                       {-2\p@ \@plus -4\p@ \@minus -4\p@}%
                       {-0.5em \@plus -0.22em \@minus -0.1em}%
                       {\normalfont\normalsize\bfseries}}
\renewcommand\section{\@startsection{section}{1}{\z@}%
                       {-8\p@ \@plus -4\p@ \@minus -4\p@}%
                       {8\p@ \@plus 4\p@ \@minus 4\p@}%
                       {\normalfont\large\bfseries\boldmath
                        \rightskip=\z@ \@plus 8em\pretolerance=10000 }}
\renewcommand\subsection{\@startsection{subsection}{2}{\z@}%
                       {-8\p@ \@plus -4\p@ \@minus -4\p@}%
                       {4\p@ \@plus 2\p@ \@minus 2\p@}%
                       {\normalfont\normalsize\bfseries}}                        

\makeatother

\title{Finding Finite Models in Multi-Sorted \\ First-Order Logic%
\thanks{This work was supported by EPSRC grant EP/K032674/1. 
Martin Suda and Andrei Voroknov were partially supported by ERC Starting Grant 2014 SYMCAR 639270. 
Andrei Voronkov was also partially supported by the Wallenberg Academy Fellowship 2014 - TheProSE.} }
\author{
Giles Reger\inst{1} \and 
 Martin Suda\inst{1} \and 
 Andrei Voronkov\inst{1,2,3}
}
\institute{
University of Manchester, Manchester, UK \and
Chalmers University of Technology, Gothenburg, Sweden \and
EasyChair
}

\begin{document}
\maketitle
\begin{abstract}
This work extends the existing MACE-style finite model finding approach to multi-sorted first-order logic. 
This existing approach iteratively assumes increasing domain sizes and encodes the related ground problem as a SAT problem. 
When moving to the multi-sorted setting each sort may have a different domain size, leading to an explosion in the search space. 
This paper focusses on methods to tame that search space. 
The key approach adds additional information to the SAT encoding to suggest which domains should be grown. 
Evaluation of an implementation of techniques in the Vampire theorem prover shows that they dramatically reduce the search space and that this is an effective approach to find finite models in multi-sorted first-order logic.
\end{abstract}

\input{introduction}
\input{prelim}

\input{singleSort}
\input{previous}

\input{multiSortFramework}
\input{sortBounds}
\input{sortInference}
\input{xmasEncoding}

\input{experiments}
\input{conclusion}

\bibliography{bib}
\bibliographystyle{cs-abbrv++}

\end{document}

%% file: introduction.tex

\section{Introduction}
\label{sec:introduction}

There have been a number of approaches looking at finding finite models for First-Order Logic (FOL), however there has not been much work on finding such models for Multi-Sorted FOL where symbols are given \emph{sorts}. We consider a model finding method, pioneered by MACE \cite{Mccune94adavis-putnam}, that encodes the search as a SAT problem. We show how this method can be modified to deal directly with multi-sorted input, rather than translating the problem to the unsorted setting, which is the most common current method.

There are two main motivations for this work. Firstly, many problems are more naturally expressed in multi-sorted FOL than in unsorted FOL (although their theoretical expressive power is equivalent). Therefore, it is useful to be able to reason in this setting and translations from multi-sorted FOL to unsorted FOL often make this reasoning harder. Secondly, MACE-style model finders can use sort information to make the SAT encoding smaller. However, as we discuss below, finding finite models of multi-sorted formulas also presents significant challenges.

The MACE-style approach, later extended in the Paradox \cite{claessen2003new} work,  involves selecting a domain size for the finite model, grounding the first-order problem with this domain and translating the resulting formulas into a SAT problem, which, if satisfied, gives a finite model of the selected size. Search for a finite model then involves considering iteratively larger domain sizes.
In the multi-sorted setting it is necessary to consider the size of each sort separately. This can be demonstrated by the following example, which is an extension of the much used \emph{Monkey Village} example \cite{inp:BlanchetteBoehmePopescuSmallbone2013,DBLP:conf/cade/ClaessenLS11}.

\begin{example}[Organised Monkey Village] \label{organisedMonkeyVillage}
Imagine a village of monkeys where each monkey owns at least two bananas. As the monkeys are well-organised, each tree contains exactly three monkeys. Monkeys are also very friendly, so they pair up to make sure they will always have a partner. We can represent this problem as follows:
\[
\begin{array}{l}
(\forall M : \textit{monkey})( {\sf owns}(M,{\sf b_1}(M)) \wedge {\sf owns}(M,{\sf b_2}(M)) \wedge {\sf b_1}(M) \neq {\sf b_2}(M)) \\
(\forall M_1,M_2 : \textit{monkey})(\forall B : \textit{banana})({\sf owns}(M_1,B) \wedge {\sf owns}(M_2,B) \rightarrow M_1 = M_2)\\
(\forall T : \textit{tree})(\exists M_1,M_2,M_3 : \textit{monkey})( (\bigwedge_{i=1}^3 {\sf sits}(M_i)=T) \wedge {\sf distinct}(M_1,M_2,M_3))\\
(\forall M_1,M_2,M_3,M_4 : \textit{monkey})(\forall T : \textit{tree})( (\bigwedge_{i=1}^4 {\sf sits}(M_i)=T) \Rightarrow \neg {\sf distinct}(M_1,M_2,M_3,M_4))\\
(\forall M : \textit{monkey})({\sf partner}(M) \neq M \wedge {\sf partner}({\sf partner}(M))=M)
\end{array}
\]
where the predicates ${\sf owns}$ associates monkeys with bananas, the functions ${\sf b_1}$ and ${\sf b_2}$ witness the existence of each monkey's minimum two bananas, the function ${\sf sits}$ maps monkeys to the tree that they sit in, the function ${\sf partner}$ associates a monkey with its partner, and the (meta-)predicate ${\sf distinct}$ is true if all of its arguments are distinct.
\end{example}

This problem requires the domain of $\textit{monkey}$ to be exactly three times larger than the domain of $\textit{tree}$,
the domain of $\textit{banana}$ to be at least twice as large as the domain of $\textit{monkey}$,
and the domain of $\textit{monkey}$ to be even. 
The main model finding effort then becomes searching for an assignment of domain sizes that satisfies this problem.
Here, the smallest such assignment is $|\textit{tree}|=2$, $|\textit{monkey}|=6$, and $|\textit{banana}|=12$.
In the general case it is necessary to try all combinations of domain sizes. 
In the worst case this will mean trying a number of assignments exponential in the number of sorts.

The techniques introduced in this paper tackle this issue by introducing a number of ways to constrain this search space. 
The main contributions can be summarised as:
\begin{enumerate}
	\item We show how the MACE-style approach (described in Sect.~\ref{sec:singleSort}) can be extended to the multi-sorted setting 
	via a novel extension of the SAT encoding (Sect.~\ref{sec:multiSortFramework}). 
	This encoding uses information from the SAT solver to guide the search through the space of domain size assignments.
	\item We use \emph{monotonic sorts} introduced in \cite{DBLP:conf/cade/ClaessenLS11} in a new way for the multi-sorted case to reduce the search space further (Sect.~\ref{sec:ourMonotonicity}).
	\item We utilise the saturation-based approach for first-order logic to detect further constraints on the search space introduced by injective and surjective functions (Sect.~\ref{sec:sortBounds}).
	 \item We present an alternative to (1), a complementary search strategy, utilising a different SAT encoding, that only needs to expand (and never shrink) the sizes of sort domains (Sect. \ref{sec:xmasEncoding}). 
\end{enumerate}
These ideas have been realised within the \vampire{} theorem prover \cite{KovacsVoronkov:CAV:Vampire:2013} and evaluated on problems taken from the TPTP and SMT-LIB benchmark suites. Our experimental evaluation (Sect.~\ref{sec:experiments}) shows that these techniques can be used to (i) improve finite model finding in the unsorted setting, (ii) effectively and efficiently find finite models in the multi-sorted setting, and (iii) detect cases where no 
models exist.



%% file: prelim.tex

\section{Preliminaries}
\label{sec:prelims}

\paragraph{Multi-Sorted First-Order Logic.} We consider a multi-sorted first-order logic with equality. 
A term is either a variable, a constant, or a function symbol applied to terms.
A literal is either a propositional symbol, a predicate applied to terms,
an equality of two terms, or a negation of either. 
Function and predicate symbols are \emph{sorted} i.e. their arguments 
(and the return value in the case of functions) have a unique \emph{sort} drawn from a finite set of sorts $S$.
We only consider well-sorted literals. There is an equality symbol per sort and equalities can only be between terms of the same sort. 
Formulas may use the standard notions of quantification and boolean connectives, but in this work we assume all formulas are \emph{clausified} using standard techniques. A \emph{clause} is a disjunction of literals where all variables are universally quantified (existentially quantified variables can be replaced by skolem functions during clausification).

\paragraph{SAT Solvers.} The technique we present later will make use of a black-box SAT solver and we assume the reader is familiar with their general properties.
We assume that a SAT solver supports \emph{solving under assumptions} \cite{DBLP:conf/sat/EenS03,DBLP:journals/entcs/EenS03}.
This means the SAT solver can be asked to search for a model of a set of clauses $N$ additionally satisfying 
a conjunction of assumption literals $A$ and is able, in case the answer is UNSAT, to provide a subset $A_0 \subseteq A$ 
of those assumptions 
which were sufficient for the unsatisfiability proof.

%% file: singleSort.tex

\section{MACE-Style Finite Model Finding in an Unsorted Setting}
\label{sec:singleSort}

We describe the finite model finding procedure in a single sorted setting. This is a variation of the approach taken by Paradox \cite{claessen2003new}. The general idea is to create, for each integer $n \geq 1$, a SAT problem that is satisfiable if the problem has a finite model of size $n$. To find a finite model we therefore iterate the approach for domain sizes $n = 1,2,3,\ldots$. 

\subsection{\DC-Models}

Let $S$ be a set of clauses. Let us fix an integer $n \geq 1$. Let $\DC=\{c_1,\ldots,c_n\}$ be a set of distinct constants not occurring in $S$, we will call the elements of $\DC$ \emph{domain constants}. We extend the language by adding the domain constants and say that an interpretation is a \emph{\DC-interpretation}, if (i) the domain of this interpretation is $\DC$ and (ii) every domain constant $c_i$ is interpreted in it by itself. Every model of $S$ that is also a \DC-interpretation will be called a \emph{\DC-model} of $S$. It is not hard to argue that, if $S$ has a model of size $n$, then it also has a \DC-model. We say that $S$ is $n$-satisfiable if it has a model of size $n$.

Let $C$ be a clause. A \emph{\DC-instance} of $C$ is a ground clause obtained by replacing every variable in $C$ by a constant in $\DC$. For example, if $p(x) \lor x = y$ is a clause and $n \geq 2$, then $p(c_1) \lor c_1 = c_2$ and $p(c_1) \lor c_1 = c_1$ are \DC-instances, while $p(c_1) \lor c_2 = c_3$ is not a \DC-instance. A clause with $k$ different variables has exactly $n^k$ \DC-instances. 

\begin{theorem}
  Let $I$ be a \DC-interpretation and $C$ a clause. Then $C$ is true in $I$ if and only if all \DC-instances of $C$ are true in $I$.\label{thm:grounding}
\end{theorem}
Let us denote by $S^*$ the set of all \DC-instances of the clauses in $S$. 
Consider an example. Let $S$ consist of three clauses
\[
p(b), \quad\quad f(a) \neq b, \quad\quad f(f(x)) = x.
\]
The smallest model of $S$ has a domain of size two. Take $n=2$, then $\DC = \{c_1,c_2\}$. By the above theorem, $S$ has a model of size two if an only if $S^*$ has a DC-model. The set $S^*$ consists of four ground clauses:

\[
  p(b), \quad\quad f(a) \neq b, \quad\quad f(f(c_1)) = c_1, \quad\quad f(f(c_2)) = c_2.
\]
Note that \DC-models are somehow similar to Herbrand models used in logic programming and resolution theorem proving, except that they are built using (domain) constants instead of all ground terms and \DC-instances instead of ground 
instances.

Theorem~\ref{thm:grounding} is not directly applicable to encode the existence of models of size $n$ as a SAT problem, because \DC-instances can contain complex terms. We will now introduce a special kind of ground atom which contains no complex subexpressions. We call a \emph{principal term} any term of the form $f(d_1,\ldots,d_m)$, where $m \geq 0$, $f$ is a function symbol, which is not a domain constant, and $d_1,\ldots,d_m$ are domain constants. In our example there are four principal terms: $a,b,f(c_1),f(c_2)$. 
A ground atom is called \emph{principal} if it either has the form $p(d_1,\ldots,d_m)$
where $m \geq 0$, $p$ is a predicate symbol different from equality and $d_1,\ldots,d_m$ are domain constants or
has the form $t = d$, where $t$ is a principal term and $d$ a domain constant. We call a \emph{principal literal} a principal atom or its negation.

\begin{theorem}
  Let $I_1,I_2$ be \DC-interpretations. If they satisfy the same principal atoms, then $I_1$ coincides with $I_2$.\label{thm:principal}
\end{theorem}
Theorem~\ref{thm:grounding} reduces $n$-satisfiability of $S$ to the existence of a \DC-interpretation of the set $S^*$ of ground clauses. Theorem~\ref{thm:principal} shows that \DC-interpretations can be identified by the set of principal atoms true in them. What we will do next is to introduce a propositional variable for every principal atom and reduce the existence of a \DC-model of $S^*$ to satisfiability of a set of clauses using only principal literals.

\subsection{The SAT Encoding}

The main step in the reduction is to transform every non-ground clause $C$ into an equivalent clause $C'$ such that \DC-instances of $C'$ consist (almost) only of principal literals. We will explain what ``almost'' means below. This transformation is known as \emph{flattening}.

\paragraph{Flattening.} A literal is called \emph{flat} if it has one of the following forms:
\begin{enumerate}
	\item $p(x_1,\ldots,x_m)$ or $\neg p(x_1,\ldots,x_m)$, where $m \geq 0$ and $p$ is a predicate symbol;
	\item $f(x_1,\ldots,x_m) = y$ or $f(x_1,\ldots,x_m) \neq y$, where $m \geq 0$ and $f$ is a function symbol, which is not a domain constant. 
	\item an equality between variables $x = y$.
\end{enumerate}
Every \DC-instance of a flat literal is either a principal literal (for the first two cases), or an equality $c_i = c_j$ between domain constants. 

To flatten clauses in $S$, we first get rid of all inequalities between variables, replacing every clause of the form $x \neq y \lor C[x]$ by the equivalent clause $C[y]$. Then we repeatedly replace every clause $C[t]$, where $t$ is not a variable and $t$ occurs as an argument to a predicate or a function symbol, by the equivalent clause $t \neq x \lor C[x]$, where $x$ is a fresh variable.

Our example clauses can be flattened as follows:
\[
p(y) \vee b \neq y, \quad\quad f(y_1) = y_2 \vee a \neq y_1 \vee b \neq y_2, \quad\quad f(y) = x \vee f(x) \neq y
\]

\paragraph{\DC-Instances.} We can now produce the \DC-instances of each flattened clause $C[x_1,\ldots,x_k]$. 
For our running example (with $n=2$) this produces the following ten \DC-instances:
\[
\begin{array}{lllll}
p(c_1) \vee b \neq c_1 &\quad\quad\quad& f(c_1) = c_1 \vee a \neq c_1 \vee b \neq c_1 &\quad\quad\quad& f(c_1) = c_1 \vee f(c_1) \neq c_1 \\
p(c_2) \vee b \neq c_2 && f(c_1) = c_2 \vee a \neq c_1 \vee b \neq c_2 && f(c_1) = c_2 \vee f(c_2) \neq c_1 \\
&& f(c_2) = c_2 \vee a \neq c_2 \vee b \neq c_2 && f(c_2) = c_2 \vee f(c_2) \neq c_2 \\
&& f(c_2) = c_1 \vee a \neq c_2 \vee b \neq c_1 && f(c_2) = c_1 \vee f(c_1) \neq c_2 \\
\end{array}
\]
Note that all literals are principal. If we treat principal atoms as propositional variables, the two leftmost clauses can be satisfied by making $b \neq c_1$ and $b \neq c_2$ both true, but this violates the assumption that $b$ should equal one of the domain constants. Additionally, the two rightmost topmost clauses can be satisfied by making $f(c_1) = c_1$ and $f(c_1)=c_2$ true but this violates the assumption that $f$ is a function. We would like to prevent both situations. To do this we introduce additional definitions. 

\paragraph{Functionality Definitions.} For each principal term $p$ and distinct domain constants $d_1,d_2$ we produce the following clause

  \[
     p \neq d_1 \vee p \neq d_2,
  \]
These clauses are satisfied by every \DC-interpretation and guarantee that all function symbols are interpreted as (partial) functions.

For our running example we introduce four new definitions:
\[
a \neq c_1 \vee a \neq c_2, \quad 
b \neq c_1 \vee b \neq c_2 , \quad 
f(c_1) \neq c_1 \vee f(c_1) \neq c_2, \quad
 f(c_2) \neq c_1 \vee f(c_2) \neq c_2
\]

\paragraph{Totality Definitions.} For each principal term $p$ we produce the following clause

\[p = c_1 \vee \ldots \vee p = c_n\] 
These clauses are satisfied by every \DC-interpretation of size $n$ and guarantee,
together with functionality axioms, that all function symbols are interpreted as total functions.

For our running example we introduce four new definitions:
\[
a = c_1 \vee a = c_2, \quad
b = c_1 \vee b = c_2, \quad
f(c_1) = c_1 \vee f(c_1) = c_2, \quad
f(c_2) = c_1 \vee f(c_2) = c_2
\]
The resulting SAT clauses have a model, meaning that the original clauses have a finite model with a domain of size 2, which can be extracted from the SAT encoding.

\paragraph{Equalities Between Variables.} Flattening can result in equalities between variables, that is, clauses of the form $C \lor x = y$. \DC-instances of such clauses can have, in addition to principal literals, equalities between domain constants $d_1 = d_2$, which are not principal literals. Since we only want to deal with principal literals, we will get rid of such equalities in an obvious way: delete clauses containing tautologies $d = d$ and delete from clauses literals $d_1 = d_2$, where $d_1$ are distinct $d_2$ domain constants. 

The following theorem underpins the SAT-based finite model building method:

\begin{theorem}
Let $S$ be a set of flat clauses and $S'$ be the set of clauses obtained from $S^*$ by removing equalities between domain constants as described above and adding all functionality and totality definitions. Then (i) all literals in $S'$ are principal and (ii) $S$ is $n$-satisfiable if and only if $S'$ is propositionally satisfiable.
\end{theorem}

\paragraph{Incrementality.} In \cite{claessen2003new} the authors describe a method for incremental finite model finding
which advocates keeping (parts of) the contents of the SAT solver when increasing $n$.
However, in previous experiments we discovered that the technique of \emph{variable and clause elimination} \cite{DBLP:conf/sat/EenB05} 
is useful at reducing the size of the SAT problem. As this is not compatible with incremental solving, 
our general approach is non-incremental. 

\subsection{Reducing the Number of Variables} 

The number of instances produced is exponential in the number of variables in a flattened clause. 
We describe two approaches that aim to reduce this number. 

\paragraph{Definition Introduction.} This reduces the size of clauses produced by flattening. Complex ground subterms are removed from clauses by introducing definitions. For example, a clause $p(f(a,b),g(f(a,b)))$ becomes $p(e_1,e_2)$ and we introduce the definition clauses $e_1 = f(a,b)$ and $e_2 = g(e_1)$, where $e_1,e_2$ are new constants. One can also introduce definitions for non-ground subterms.

\paragraph{Clause Splitting.} Clauses with $k$ variables are split into subclauses having less than $k$ variables each. New predicate symbols applied to the shared variables are then added to join the subclauses. For example, the clause $p(x,y) \vee q(y,z)$ with three variables is replaced by the two clauses $p(x,y) \vee s(y)$ and $\neg s(y) \vee q(y,z)$ where $s$ is a new predicate symbol. These new clauses have two variables each. For large domain sizes splitting can drastically reduce the size of the resulting propositional problem. This was first used for finite model finding by Gandalf \cite{DBLP:journals/jar/Tammet97} and later in Eground \cite{DBLP:conf/flairs/Schulz02} for EPR problems.

\subsection{Symmetry Breaking}

The SAT problem produced above can contain many symmetries. For example, every permutation of $\DC$ applied to a \DC-model will give a \DC-model, and there are $n!$ such permutations. 
We can (partially) break these symmetries as follows. Firstly, if the input contains constants $a_1,\ldots,a_l$ we can add the clauses 
\[
a_i \neq c_m \vee a_1 = c_{m-1} \vee \ldots \vee a_{i-1} = c_{m-1} 
\]
for $1 < i \le l$ and $ 1< m \le n$, 
where we have arbitrarily ordered the constants and captured the constraint that if the $i$-th constant is equal to a domain element then some earlier constant must be equal to the next smallest domain element. 
Secondly, we can tell the SAT solver about this order on constants by adding the clauses
		\[
		a_i = c_1 \vee \ldots \vee a_i = c_i
		\]
for $i \le {\sf min}(m,n)$, which captures the constraint that the $i$-th constant must be equal to one of the first $i$-th domain elements. 
If $1 < m < n$ then we can also use principal terms other than constants in the second case, but not in the first.

\subsection{Determining Unsatisfiability}

If it is possible to detect the \emph{maximum} domain size then it is possible to show there is no model for a formula if all domain sizes up to, and including, this maximum size have been explored. There are two straightforward ways to detect maximum domain sizes. 
Firstly, we can look for axioms such as $(\forall x)(x = a \vee x =b)$ and $(\forall x)(\forall y)(\forall z)(x = y \vee x=z \vee z=y)$. Both indicate that the problem has a maximum domain size of 2. 
Secondly, we can look for so-called EPR problems that only use constant function symbols, in this case, the domain size is bounded by the number of constants.
%

%% file: previous.tex

\section{Previous Work in the Multi-Sorted Setting}
\label{sec:previousMulti}

We review previous work related to finite model finding for multi-sorted FOL.

\paragraph{Translating Sorts Away.} \label{sec:translation}

One approach to dealing with multi-sorted FOL is to translate the sorts away. 
We discuss two well-known translations, see \cite{inp:BlanchetteBoehmePopescuSmallbone2013} for further discussions of such translations.

\textit{Sort Predicates.} One can \emph{guard} the use of sorted variables by a \emph{sort predicate} that indicates whether a variable is of that sort. This predicate can be set to false in a model for all constants not of the appropriate sort. For example, the last formula in the Organised Monkey Village problem can be rewritten using the sort predicate ${\sf isMonkey}$.
\[
(\forall M)({\sf isMonkey}(M) \rightarrow {\sf partner}(M) \neq M \wedge {\sf partner}({\sf partner}(M))=M)
\]
One also needs to add additional axioms that say that sorts are non-empty and that functions return the expected sort. For the \textit{monkey} sort we need to add
\[
(\exists M)({\sf isMonkey}(M)) \quad\quad (\forall M)({\sf isMonkey}({\sf partner}(M))).
\]

\textit{Sort Functions or Tags.} One can \emph{tag} all values of a sort using a \emph{sort function} for that sort. The idea is that in a model the function can map all constants (of any sort) to a constant of the given sort. For example, the last formula from the Organised Monkey Village problem can be rewritten using ${\sf f_m}$ as a sort function for \textit{monkey}:
\[
(\forall M)({\sf f_m}({\sf partner}({\sf f_m}(M))) \neq {\sf f_m}(M) \wedge {\sf f_m}({\sf partner}({\sf f_m}({\sf partner}({\sf f_m}(M)))))={\sf f_m}(M))
\]
The authors of \cite{inp:BlanchetteBoehmePopescuSmallbone2013} suggest conditions that allow certain sort predicates and functions to be omitted. However, their arguments relate to resolution proofs and do not apply here.

\paragraph{Sorting it Out with Monotonicity.} \label{sec:monotonicity}

In \cite{DBLP:conf/cade/ClaessenLS11} Claessen et al. introduce a monotonicity analysis and show how it can help translate multi-sorted formulas to unsorted ones by only applying the above translations to non-monotonic sorts. 
A sort $\tau$ is monotonic for a multi-sorted FOL formula $\phi$ if for any model of $\phi$ one can add an element to the domain of $\tau$ to produce another model of $\phi$.
%
For example, in the Organised Monkey Village example the \textit{banana} sort is monotonic as we can add more bananas once we have enough. However, \textit{monkey} and \textit{tree} are not monotonic as increasing either requires more trees, monkeys and bananas.

In \cite{DBLP:conf/cade/ClaessenLS11} they observe that if there is no positive equality between elements of a sort then a new domain constant can be added and made to behave like an existing domain constant and there is no way to detect this i.e. positive equalities are required to bound a sort. They refine this notion further by noting that a positive equality can be guarded by a predicate, if that predicate can be forced to be true for all new domain elements. They introduce a calculus and associated SAT encoding capturing these ideas that can be used to detect monotonic sorts, which we use in our work. 
%
%
\paragraph{Using a Theory of Sort Cardinalities.}
In the single-sorted setting there is a family of techniques called SEM-style after the SEM model finder \cite{DBLP:conf/ijcai/ZhangZ95}  based on constraint satisfaction methods. 
There exists a technique in this direction for the multi-sorted setting implemented in the CVC4 SMT solver \cite{DBLP:conf/cav/ReynoldsTGK13}. 
The idea behind this approach is to introduce a theory of sort cardinality constraints and to incorporate this theory into the standard SMT solver structure. 
Briefly, this approach introduces cardinality constraints (upper bounds) for sorts and searches for a set of constraints that is consistent with the axioms. To check a cardinality constraint $k$ for sort $s$, a congruence relation is built for $s$-terms and an attempt made to merge congruence classes so that there are at most $k$. 
Cardinality constraints are then increased if found to be inconsistent. 
Quantified formulas are then instantiated with representative constants from the equivalence classes.

%% file: multiSortFramework.tex

\section{A Framework for the Multi-Sorted Setting}
\label{sec:multiSortFramework}

In this section we introduce our framework able to build models of multi-sorted formulas directly, in contrast to translating the sorts away. The key challenge is dealing with a large and growing search space of domain sizes.

\subsection{Using Sorts in the SAT Encoding}

The SAT encoding in Sect.~\ref{sec:singleSort} can be updated to become sort-aware. First, instead of the domain size $n$ we use finite domain sizes $n_s$ for every sort $s$. Second, instead of considering $\DC=\{c_1,\ldots,c_n\}$, we consider domains for each sort $\DC_s=\{c_1,\ldots,c_{n_s}\}$. We can now define $n$ as a function (called \emph{domain size assignment}) mapping each sort $s$ to $n_s$ and likewise, define $\DC$ as the function mapping each sort $s$ to $\DC_s$. After that we can speak about \DC-models and $n$-satisfiability in the multi-sorted case. 

All the definitions for the one-sorted case are modified to respect sorts.
This means, in particular, that in a $\DC$-instance of a clause a variable of a sort $s$ can only be replaced by a domain constant in $\DC_s$. 
For example, for the Organised Monkey Village problem (see page~\pageref{organisedMonkeyVillage}) we could consider the domain size assignment $n$ such that $n_\mathit{tree}=1$, $n_\mathit{monkey}=2$ and $n_\mathit{banana}=2$. The first formula in this description can be split into three clauses, the first of which would be flattened as 
$
{\sf owns}(M,x) \vee {\sf b_1}(M) \neq x
$, 
which would have the two \DC-instances 
$
{\sf owns}(c_1,c_1) \vee {\sf b_1}(c_1) \neq c_1$
and
${\sf owns}(c_1,c_2) \vee {\sf b_1}(c_1) \neq c_2$. 
We can use $c_1$ for both monkeys and bananas here as monkeys and bananas are never compared. For this reason we can also break symmetries on a per-sort basis.

Once we have updated the SAT encoding, finite model finding can then proceed as before where we construct the SAT problem for the current domain size assignment, check for satisfiability, and then either return a model or repeat the process with an updated domain size assignment. The problem then becomes how to generate the next domain size assignment to try.

\subsection{A Search Strategy}

We will view the search space of domain size assignments as an infinite directed graph whose nodes are domain size assignments and the children of an assignment are all the nodes that have exactly one domain size that is one larger.
Thus, the number of children of every node is the number of sorts. 
A child of a node $n$ having a larger domain size than $n$ for a sort $s$ is called the \emph{$s$-child} of $n$. The \emph{$s$-descendant} relation is the transitive closure of the $s$-child relation. 

A \emph{search strategy} will explore this graph node by node in such a way that a node is always visited before its children. For each node $n$ that we visit, we can either check $n$-satisfiability or \emph{ignore} this node. To decide whether a node can be ignored, we will maintain a set of \emph{constraints}. 
Abstractly, a constraint is a predicate on domain size assignments and nodes that do not satisfy the current set of constraints will be ignored. 
Concretely, we will use a language of (boolean combinations of) arithmetical comparison literals such as $|s| < b$, $|s| \leq b$, \ldots 
to represent the constraints. Here $b$ stands for a concrete integer and $|s|$ is 
a symbolic placeholder variable for the ``intended size'' 
of the domain of sort $s$.
The semantics of the this representation is the obvious one.

We will work with a queue $Q$ of nodes and a set $\CS$ of \emph{constraints}. Initially, $Q$ consists of a single node assigning 1 to all sorts and $\CS$ is empty. We then repeat the following steps:

\begin{enumerate}
\item If $Q$ is empty, return ``unsatisfiable''.
\item Remove the node $q$ from the front of $Q$. Do nothing if $q$ was visited before at this step. Otherwise, continue with the following steps.
\item If $q$ satisfies all constraints in $\CS$, perform finite model finding for $q$, terminating if a model is found. In variations of this algorithm considered later, we can add some constraints to $\CS$ at this step: these constraints will be obtained by analyzing the proof of $q$-unsatisfiability.
\item Add to $Q$ all children of $q$.
\end{enumerate}


We will now introduce an important notion helping us to prevent exploring large parts of the search space. 
A constraint is said to have the \emph{$s$-beam} property at a node $n$, if all $s$-descendants of $n$ violate this 
constraint. For example, the constraint $|s| < 3$ has the $s$-beam property at any node $q$ having $q_s=2$. We can generalize this notion to more than one sort. 

With this notion we can improve step 4 of the algorithm as follows:

\begin{enumerate}
\item[$4.'$] If there is a constraint in $\CS$ having an $s$-beam property at the $s$-child $n$ of $q$, if $n$ violates this constraint, add to $Q$ all children of $q$ apart from $n$.
\end{enumerate}
For example, if we have the constraint $|s| < 3$ and $q_s=2$, this constraint will prevent us from considering the $s$-child $n$ of $q$ having $n_s=3$.

%


\label{page:heuristic} As a small refinement, we introduce a heuristic for deciding which node in the queue to consider next, rather than processing them in the first-in-first-out order. The idea is to estimate how difficult a domain size assignment is to check
and to prioritise exploration of the easier parts of the search space. 
Under this variation, $Q$ is a priority queue ordered by some size measure
of the corresponding SAT encoding (in the experiment, we measured size in the number of clauses).
This setup is complete, as long as this size grows strictly from a parent to its child (which is trivially satisfied for number of clauses).



\subsection{Encoding the Search Problem}


We now show how an extension of the SAT encoding can be used to produce constraints and therefore indicate areas of the search space that should be avoided.
This is done by marking certain clauses of the encoding with certain special variables and using the mechanism for solving under assumptions
to detect which of these clauses were actually used in the unsatisfiability proof. 
We will design the names of these special marking variables in such a way
that the detected set of used assumptions will immediately correspond to a (disjunctive) constraint.

Let us assume we are encoding for the domain size assignment $n$. 
For each sort $s$ we introduce two new propositional variables $``|s|>n_s"$ and $``|s|<n_s"$,
which can be understood as stating that the intended size of the domain of $s$ should be larger,
respectively smaller, than current $n_s$. The marking of clauses is now done as follows.

The totality definition for each principal term $p$ becomes
\[
p = c_1 \vee \ldots \vee p = c_{n_s} \vee ``|s|>n_s"
\]
i.e. either the principal term equals one of the domain constants or the domain is currently too small. 
\DC-instances can be similarly updated. Let $C$ be a \DC-instance and let ${\sf sorts}(C)$ be the set of sorts of variables occurring in $C$.
We replace $C$ by
\[ \textstyle
C \vee \bigvee_{s \in{\sf sorts}(C)} ``|s|<n_s"
\]
i.e. either the \DC-instance holds or the domain is too large.

We then attempt to solve the updated SAT problem under the assumptions
\[ \textstyle
A = \bigwedge_{s \in S} (\neg ``|s|>n_s") \wedge (\neg ``|s|<n_s")
\]
i.e. we assume that we are using the correct domain sizes.
These added assumptions ensure that the logical meaning of the updated encoding
is exactly the same as before. However, in the unsatisfiable case
the solver now returns a subset $A_0 \subseteq A$ of the assumptions 
that were sufficient to establish unsatisfiability. Equivalently,
$\neg A_0$ is a conflict clause over the marking variables implied by the encoded problem.
This clause can now be understood as the newly derived constraint. 
We just need to interpret the marking variables in their ``unquoted'' form,
i.e., as arithmetic comparison literals.

\begin{figure}[t]
\centering
\begin{tikzpicture}[scale=0.9]

  \draw[style=help lines,step=0.5cm] (0,0) grid (1.9,3.4);
  \draw[->] (0,0) -- (2,0) node[right] {$\mathit{tree}$};
  \draw[->] (0,0) -- (0,3.5) node[above] {$\mathit{monkey}$};

\foreach \x/\xtext in {1,2,3}
    \draw[shift={(\x/2-0.25,0)}] (0pt,2pt) -- (0pt,-2pt) node[below] {$\xtext$};

  \foreach \y/\ytext in {1,2,3,4,5,6,7}
    \draw[shift={(0,\y/2-0.25)}] (2pt,0pt) -- (-2pt,0pt) node[left] {$\ytext$};

  \draw (0.25,0.25) circle (0.05cm) node[label={[red,label distance=-0.21cm]135:{\tiny 1}}]{};
  \draw (0.25,0.75) circle (0.05cm) node[label={[red,label distance=-0.21cm]135:{\tiny 2}}]{};
  \draw (0.25,1.25) circle (0.05cm) node[label={[red,label distance=-0.21cm]135:{\tiny 3}}]{};  
  \draw (0.25,1.75) circle (0.05cm) node[label={[red,label distance=-0.21cm]135:{\tiny 4}}]{};
  \draw (0.75,1.75) circle (0.05cm) node[label={[red,label distance=-0.15cm]90:{\tiny 5-7}}]{};
  \draw (0.75,2.25) circle (0.05cm) node[label={[red,label distance=-0.21cm]135:{\tiny 8}}]{};
  \draw (0.75,2.75) circle (0.05cm) node[label={[red,label distance=-0.15cm]90:{\tiny 9-17}}]{};
\draw[->,draw,red](0.25,0.25) -- (1.75,0.25){}; 
\draw[->,draw,red](0.25,0.75) -- (1.75,0.75){};
\draw[->,draw,red](0.25,1.25) -- (1.75,1.25){};
\draw[->,draw,red](0.25,1.75) -- (0.25,3.25){};
\draw[->,draw,red](0.75,1.75) -- (1.75,1.75){};
\draw[->,draw,red](0.75,2.25) -- (1.75,2.25){};
\draw[->,draw,red](0.75,2.75) -- (1.75,2.75){};

\end{tikzpicture}
\hspace{1cm}
\begin{tikzpicture}[scale=0.9]

  \draw[style=help lines,step=0.5cm] (0,0) grid (5.9,3.4);
  \draw[->] (0,0) -- (6,0) node[right] {$\mathit{banana}$};
  \draw[->] (0,0) -- (0,3.5) node[above] {$\mathit{monkey}$};

\foreach \x/\xtext in {1,2,3,4,5,6,7,8,9,10,11,12}
    \draw[shift={(\x/2-0.25,0)}] (0pt,2pt) -- (0pt,-2pt) node[below] {$\xtext$};

  \foreach \y/\ytext in {1,2,3,4,5,6,7}
    \draw[shift={(0,\y/2-0.25)}] (2pt,0pt) -- (-2pt,0pt) node[left] {$\ytext$};

  \draw (0.25,0.25) circle (0.05cm) node[label={[red,label distance=-0.21cm]135:{\tiny 1}}]{};
  \draw (0.25,0.75) circle (0.05cm) node[label={[red,label distance=-0.21cm]135:{\tiny 2}}]{};
  \draw (0.25,1.25) circle (0.05cm) node[label={[red,label distance=-0.21cm]135:{\tiny 3}}]{};
  \draw (0.25,1.75) circle (0.05cm) node[label={[red,label distance=-0.15cm]270:{\tiny 4-5}}]{};
  \draw (0.75,1.75) circle (0.05cm) node[label={[red,label distance=-0.21cm]135:{\tiny 6}}]{};
  \draw (1.25,1.75) circle (0.05cm) node[label={[red,label distance=-0.21cm]135:{\tiny 7}}]{};
  \draw (1.25,2.25) circle (0.05cm) node[label={[red,label distance=-0.21cm]135:{\tiny 8}}]{};
  \draw (1.25,2.75) circle (0.05cm) node[label={[red,label distance=-0.21cm]135:{\tiny 9}}]{};
\draw (1.75,2.75) circle (0.05cm) node[label={[red,label distance=-0.24cm]135:{\tiny 10}}]{};
\draw (2.25,2.75) circle (0.05cm) node[label={[red,label distance=-0.24cm]135:{\tiny 11}}]{};
\draw (2.75,2.75) circle (0.05cm) node[label={[red,label distance=-0.24cm]135:{\tiny 12}}]{};
\draw (3.25,2.75) circle (0.05cm) node[label={[red,label distance=-0.24cm]135:{\tiny 13}}]{};
\draw (3.75,2.75) circle (0.05cm) node[label={[red,label distance=-0.24cm]135:{\tiny 14}}]{};
\draw (4.25,2.75) circle (0.05cm) node[label={[red,label distance=-0.24cm]135:{\tiny 15}}]{};
\draw (4.75,2.75) circle (0.05cm) node[label={[red,label distance=-0.24cm]135:{\tiny 16}}]{};
\draw (5.25,2.75) circle (0.05cm) node[label={[red,label distance=-0.24cm]135:{\tiny 17}}]{};
\draw (5.75,2.75) circle (0.05cm) node[label={[red,label distance=-0.24cm]135:{\tiny 18}}]{};

\draw[->,draw,red](0.25,0.25) -- (5.75,0.25){}; 
\draw[->,draw,red](0.25,0.75) -- (5.75,0.75){}; 
\draw[->,draw,red](0.25,1.25) -- (5.75,1.25){}; 
\draw[->,draw,red](0.25,1.75) -- (0.25,3.25){};
\draw[->,draw,red](0.75,1.75) -- (0.75,3.25){};

\draw[->,draw,red](1.25,1.75) -- (5.75,1.75){};
\draw[->,draw,red](1.25,2.25) -- (5.75,2.25){};

\draw[->,draw,red](1.25,2.75) -- (1.25,3.25){};
\draw[->,draw,red](1.75,2.75) -- (1.75,3.25){};
\draw[->,draw,red](2.25,2.75) -- (2.25,3.25){};
\draw[->,draw,red](2.75,2.75) -- (2.75,3.25){};
\draw[->,draw,red](3.25,2.75) -- (3.25,3.25){};
\draw[->,draw,red](3.75,2.75) -- (3.75,3.25){};
\draw[->,draw,red](4.25,2.75) -- (4.25,3.25){};
\draw[->,draw,red](4.75,2.75) -- (4.75,3.25){};
\draw[->,draw,red](5.25,2.75) -- (5.25,3.25){};
\end{tikzpicture}\vspace{-1em}
\caption{Finding a finite model for the Organised Monkey Village problem.\label{sec:exampleOMV}}
\end{figure}
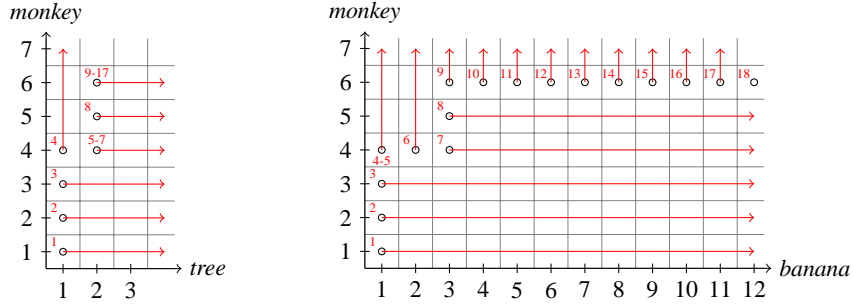

The argument why this interpretation is correct is best done with the set $A_0$,
which contains the marking variables negated. It consists of two main observations:
\begin{enumerate}
\item
	If $A_0$ contains $\neg ``|s|>n_s"$, the unsatisfiability relies on a totality clause for the sort $s$.
	Because a totality clause gets logically stronger when the domain size is decreased,
	essentially the same unsatisfiability proof could be repeated for the domain size $|s|$
	smaller or equal to the current $n_s$ (given the other conditions from $A_0$).
\item
	If $A_0$ contains $\neg ``|s|<n_s"$, the unsatisfiability relies on a \DC-instance with a variable of sort $s$.
	Because we only add more instances of a clause if a domain size is increased,
	the same unsatisfiability proof would also work for the domain size $|s|$
	greater or equal to the current $n_s$.
\end{enumerate}
Thus at least one of the (atomic) constraints represented by the literals in $A_0$ must be violated by a
domain size assignment, if we want to have a chance of finding a model.




\subsection{An Example}

Let us consider the Organised Monkey Village example (page~\pageref{organisedMonkeyVillage}). Running the initial search strategy on this problem requires checking 2,661 different domain sort assignments. Using the encoding described above means that only 18 assignments are tried. The search carried out by our approach is illustrated in Fig.~\ref{sec:exampleOMV}. We give two projections of the 3-dimensional search space and the arrows show the parts of the search space ruled out by $s$-beam constraints. On each step the constraints rule out all but one neighbour, meaning that we take a direct path to the solution through the search space.


\subsection{Using Monotonicity} \label{sec:ourMonotonicity}

In our framework we can use the notion of monotonicity (see Sect.~\ref{sec:monotonicity}) in two ways.

\begin{itemize}
\item 
\textit{Collapsing Monotonic Sorts.} All monotonic sorts can be collapsed into a single sort as this sort can grow to the size of the largest monotonic sort. 
%
It is never safe to collapse a monotonic sort into a non-monotonic one as the monotonic sort may depend on the non-monotonic one. For example, whilst \textit{banana} is a monotonic sort it must always be twice as large as the non-monotonic sort \textit{monkey}.

\item 
\textit{Refining the Search Encoding.} If a sort $s$ is monotonic then if no model exists for domain size $n_s$
then no model can exist where $n_s$ is smaller. We reflect this in our encoding 
by not marking \DC-instances with marking variables for monotonic sorts. This leads to a derivation
of potentially stronger constraints.
\end{itemize}

%% file: sortBounds.tex

\section{Detecting Constraints Between Sorts}
\label{sec:sortBounds}

In this section we discuss how properties of functions between sorts can be used to further constrain the domain size search space. 
Consider the following set of formulas
\[
\begin{array}{ll}
{\sf distinct}(a_1,a_2,a_3,a_4,a_5) \quad\quad&
(\forall x : s_1)( f(x) \neq b)\\
(\forall x,y : s_1)( f(x)=f(y) \rightarrow x=y) \quad\quad&
(\forall x,y : s_2)( g(x)=g(y) \rightarrow x=y)\\
\end{array}
\]
where $a_i$ are constants of sort $s_1$, $b$ is a constant of sort $s_2$,
$f : s_1 \rightarrow s_2$, and $g : s_2 \rightarrow s_3$. 
The previous approach would try increasing the size of each sort by 1, 
discovering that one sort must grow at each step. 
However, we can see from the bottom two formulas that $f$ and $g$ are injective 
and therefore that $|s_1| \le |s_2| \le |s_3|$, furthermore, 
the second formula tells us that $f$ is non-surjective and therefore that $|s_1| < |s_2|$. 
Using these constraints we can immediately discount 9 of the 15 domain size assignments considered without them.

To find constraints between the sizes of sorts $s_1$ and $s_2$ we look for four cases:
\begin{enumerate}
	\item If a function $f : s_1 \rightarrow s_2$ is injective, then $|s_1| \le |s_2|$.
	\item If a function $f : s_1 \rightarrow s_2$ is injective and non-surjective, then $|s_1| < |s_2|$.
	\item If a function $f : s_1 \rightarrow s_2$ is surjective, then $|s_1| \ge |s_2|$.
	\item If a function $f : s_1 \rightarrow s_2$ is surjective and non-injective, then $|s_1| > |s_2|$.
\end{enumerate}
These constraints can be added to the constraints used in the search described in Sect.~\ref{sec:multiSortFramework}. 

%
%
Our method for detecting bounds was inspired by Infinox \cite{DBLP:journals/jar/ClaessenL11}, a method for showing no finite model can exist for unsorted FOL formulas if there is a strict bound within a sort. We detect bounds by attempting to prove properties of functions \emph{between} sorts. 
%
For a unary function $f : s_1 \rightarrow s_2$ occurring in the problem we can simply make a claim such as
\[
(\forall x : s_1)(\forall y : s_2)( f(x) = f(y) \rightarrow x=y) \wedge (\exists y : s_2)(\forall x : s_1)(f(x) \neq y)
\]
for each case (this is case (2) above), and then ask whether this claim follows from the axioms of the input problem. 
For non-unary functions it is necessary to existentially quantify over one of the arguments, details of how to do this can be found in \cite{DBLP:journals/jar/ClaessenL11}.


To check each claim $C$ we could use standard techniques to check $A \models C$ where $A$ are the input axioms. Any black box solver could be used for this. 
However, doing this on a per-claim basis is inefficient and we implement an optimisation of Vampire's saturation loop to establish multiple claims in a single proof attempt. 
Recall that the saturation loop will search for consequences of its input. Therefore, we saturate $A \cup \{ C_i \rightarrow l_i \}$ where $l_i$ is a fresh propositional symbol labelling claim $C_i$. If the unit $l_i$ is derived 
 then we can conclude that the claim $C_i$ is a consequence of $A$. 
This approach was inspired by the consequence elimination mode of Vampire \cite{HoderKovacsVoronkov:MICAI:InvariantGeneration:2011} (see this work for technical details).

%

%% file: sortInference.tex

\section{Getting More Sorts}
\label{sec:sortInference}

Previously we have seen how sort information can be used to reduce the size of the SAT encoding by only growing the domain sizes of sorts that need to be grown. In this section we recall a technique first described in \cite{DBLP:conf/cade/ClaessenLS11} for inferring new sorts and explain how these new sorts can be useful.

\paragraph{Inferring Subsorts.}

Consider the Organised Monkey Village example. The $\textit{monkey}$ sort can be split into three separate subsorts as there are three parts (assigning bananas to monkeys, assigning monkeys to trees and assigning monkeys to their partners) where the signatures do not overlap. Abstractly, we can use different monkeys in these different places as they do not interact -- later we will see why this is useful. 
To infer such subsorts we can use the standard union-find method on positions in the signature. 

\paragraph{Using Inferred Subsorts.}

 
Claessen et al.  \cite{DBLP:conf/cade/ClaessenLS11} describe two uses for inferred subsorts:

-- \noindent\emph{Removing Instances.}  If a subsort $\tau$ is monotonic and all function symbols with the return sort $\tau$ are constants, then we can bound the subsort by the number of constants. It is easy to argue that any ground clauses (instances, totality or functionality) for a domain constant larger than the bound of the subsort can be omitted as they will necessarily be equivalent to an existing clause. This helps reduce the size of the SAT encoding.

-- \noindent\emph{Symmetry breaking.} For the same reasons that symmetry breaking can occur per sort, symmetry breaking can now occur per inferred subsort. This is safe due to the above observation that values for different subsorts will never be compared.

\paragraph{Making Subsorts Proper Sorts.}

Proper sorts and inferred subsorts are treated differently as we only grow the sizes of proper sorts. If an inferred subsort is not bounded as described above then it is forced to grow to the same size as its parent sort.
To understand why this can be problematic consider the FOL formula
\[
{\sf distinct}(a_1,\ldots,a_{50}) \wedge (\forall x)(f(f(f(f(f(f(f(f(f(f(x)))))))))) \neq x)
\]
which has an overall finite model size of 50. Establishing this finite model requires a SAT problem consisting of 1,187,577 clauses. 
However, there are two subsorts: that of the constants $a_1$ to $a_{50}$ and that of $f$. The second subsort is monotonic and does not need to grow beyond size 3. If this had been declared as a separate sort then the required SAT encoding would only consist of 125,236 clauses.

It is only safe to treat an inferred subsort as a proper sort if we can translate any resulting model into one where elements of the inferred subsort belong to the original sort. This is possible when (i) the inferred subsort is monotonic, and (ii) the size of the inferred subsort is not larger than the size of its parent sort. To ensure (ii) we add constraints to the search strategy in the same way as for sort bounds detected previously.

%% file: xmasEncoding.tex

\section{An Alternative Growing Search}
\label{sec:xmasEncoding}

The previous search strategy considers each domain size assignment separately
 (we therefore refer to it as a \emph{pointwise} encoding). 
However, we can modify the encoding so that it captures 
the current assignment \emph{and all smaller ones} at the same time.
Thus we no longer talk of a domain size but rather of a domain size upper \emph{bound}, as the parameter of the encoding.
These bounds never need to shrink and thus grow monotonically for each sort.
We call this encoding a \emph{contour} encoding as we can think of it drawing a contour around the explored part and growing this outwards.

This alternative encoding works as follows. For each sort $s$ with domain size bound $n_s$
we introduce $n_s$ propositional variables $\mathit{bound}_s(1)$ to $\mathit{bound}_s(n_s)$. 
Then instead of single totality constraint 
for each principal term $p$ we introduce all totality constraints for domain sizes up to $n_s$ guarded by the appropriate bound i.e.
\[
p = c_1 \vee \mathit{bound}_s(1),\quad
\ldots,\quad
p =c_1 \vee \ldots \vee p = c_{n_s} \vee \mathit{bound}_s(n_s)
\]
We guard \DC-instances of clauses with negations of these guards in the following way.
For each sort $s$ let $s_\mathit{max}$ be 
the index of the largest domain constant in this instance used to replace a variable of sort $s$.
Then if $s_\mathit{max}$ is defined, i.e. there is at least one such variable, and $s_\mathit{max} > 1$
we guard the instance with a literal $\neg \mathit{bound}_s(s_\mathit{max}-1)$. For example,
given a function symbol $f : s_1 \rightarrow s_2$, a constant $b : s_2$,
and a flattened clause $f(x) \neq y \vee b \neq y$, its \DC-instance $f(c_3) \neq c_1 \vee b \neq c_1$ would be guarded as $f(c_3)
\neq c_1 \vee b \neq c_1 \lor \neg \mathit{bound}_{s_1}(2)$.

In this encoding the SAT solver can satisfy the clauses for a domain size smaller than $n_s$
i.e. if it can satisfy a stricter totality constraint then it can effectively ignore some of the instances.
%
As a further variation, if a sort is monotonic then we do not need to consider the possibility 
that a sort is smaller than its current bound. Therefore, we only need the largest totality constraint and do not need constraints on instances.

In a similar way as before, we solve the problem under the assumptions that the sort sizes are big enough i.e.
\[ \textstyle
A = \bigwedge_{s \in S} \neg \mathit{bound}_s(n_s).
\]
If this is shown unsatisfiable the subset of assumptions $A_0$ will suggest the sorts that could be grown; 
growing a sort not mentioned in $A_0$ would allow the same proof of unsatisfiability to be produced.
If $A_0$ is empty, the SAT-solver has shown that the given first-order formula is unsatisfiable.
Otherwise, we can either arbitrarily select a sort to grow out of the ones mentioned in $A_0$.
This approach is significantly different from the previous approach as now we only consider one next domain size assignment. However, the SAT problems may be considerably harder to solve as the SAT solver is now considering a much larger set of models. In essence, each new SAT problem contains all the previous ones as sub-problems.


Finally, if the SAT problem is satisfiable then the actual size of a sort $s$ is given by its smallest totality constraint that is ``enabled'';
more precisely, by the smallest $i$ such that $\mathit{bound}_s(i)$ is false in the computed model.

%% file: experiments.tex

\section{Experimental Evaluation}
\label{sec:experiments}

In this section we evaluate the different techniques for finite model finding in multi-sorted FOL described in this paper and compare our approach to other tools. 

\paragraph{Experimental Setup.}

We considered two sets of problems. From the TPTP \cite{TPTP} library (version 6.3.0) we took unsorted problems in the FOF or CNF format.
From the SMT-LIB library \cite{BarST-SMTLIB} we took problems from the UF (Uninterpreted Functions) logic. 
Experiments were run on the StarExec cluster \cite{starexec},
whose nodes are equipped with Intel Xeon 2.4GHz processors and 128 GB of memory. 
For each experiment we will report the number of problems solved with the time limit of 60 seconds.

On satisfiable problems we compare our implementation 
with version 3.0 of Paradox \cite{claessen2003new} and version 1.4 of CVC4 \cite{DBLP:conf/cav/ReynoldsTGK13};
Paradox does not establish unsatisfiability and CVC4 runs more than a finite model finding approach,
making a comparison on unsatisfiable problems difficult. 
On the TPTP problems, we also compare to version 2.0 of iProver \cite{iprover}.
The techniques described in this paper were implemented in \Vampire{}.

\paragraph{Adding Sorts to Unsorted Problems.}

\begin{table}[t]
\caption{Experimental Results for Unsorted problems.\label{tab:results:unsorted}}
\centering
\begin{tabular}{l|ccc|ccc}
&&& \multicolumn{3}{c}{\Vampire} \\
					 	& CVC4	& Paradox & iProver	& Ignore	& Use 	& Expand  \\
\hline
FOF+CNF: sat 			& 1181		&	1444 & 1348 &     1421	& 1463& {\bf 1503}	\\
FOF+CNF: unsat 		    & -	        &	-    & 1337	&     1400	&1604 & {\bf 1628}	\\
\end{tabular}
\end{table}

Our first experiment considers the effect of sort inference on unsorted problems. We consider three settings: (i) inferred subsorts are ignored, (ii) inferred subsorts are used to reduce the problem size and break symmetries only, and (iii) inferred subsorts are expanded to proper sorts where possible. 
Table~\ref{tab:results:unsorted} presents the results. This shows that sort information can be used to solve more problems. For satisfiable problems the best \Vampire{} strategy solves more problems than CVC4, Paradox or iProver. For both satisfiable and unsatisfiable problems, expanding subsorts into proper sorts and treating the problems as multi-sorted problems helps solve the most problems.
%
We note that 4 problems found unsatisfiable using this approach could not be solved by any other technique in \Vampire{}, this is significant as \Vampire{} is one of the best theorem provers available for such problems.




\paragraph{Removing Sorts from Sorted Problems.}

Next we consider the translation techniques described in Sect.~\ref{sec:translation} applied to multi-sorted problems. Table~\ref{tab:results:translation} shows the results of running variations of these translations on the multi-sorted UF problems described above. Plain applies the translation to the whole problem, ignoring subsorts in the result. With Monotonicity only non-monotonic sorts are translated and with Subsorts the resulting problem is solved using inferred subsorts. Both adds both variations.

These results show that, for these problems, sort predicates are more useful and that the techniques of monotonicity detection and subsort inference are useful in improving the translation and reasoning with it.

\begin{table}[t]
\caption{Experimental Results for Translations from Multi-Sorted to Unsorted.\label{tab:results:translation}}
\centering
\begin{tabular}{l|cccc|cccc}
					 	& \multicolumn{4}{c}{Sort Predicates} &  \multicolumn{4}{c}{Sort Functions} \\
					 	& Plain & Monotonicity & Subsorts & Both & Plain & Monotonicity & Subsorts & Both \\
\hline
UF: sat 			& 813& 810&872 &{\bf 874} &710  &771 &834 & 873 \\
UF: unsat 			& 101& 112&221 &{\bf 232} &67 & 67&171 & 171 \\
\end{tabular}
\end{table}

\paragraph{Finding Models of Multi-sorted Problems.}

Finally, we consider our framework for reasoning with multi-sorted problems directly. 
Table~\ref{tab:results:sorted} gives the results for CVC4 and eight variations of the techniques presented in this paper (we use only CVC4 since Paradox does not work on sorted problems). At the top level these are split into the Pointwise and Contour encodings and a version where no constraints were added. Then Expand refers to subsort expansion (Sect.~\ref{sec:sortInference}),  Collapse refers to collapsing monotonic sorts together (Sect.~\ref{sec:ourMonotonicity}), and Bounds refers to sort bound extraction (Sect.~\ref{sec:sortBounds}). 
These results can also be compared to Table~\ref{tab:results:translation} as the problems are the same.

The three main conclusions from this information are (i) overall the approach taken in this paper is able to solve more problems than the approach taken by CVC4, (ii) collapsing monotonic sorts is very useful, 
and (iii) including the search problem as part of the SAT encoding is vital. 
Bracketed numbers show unique problems solved by an approach. This shows that although CVC4 solves fewer problems it does solve some uniquely.
The contour encoding was generally more successful, however in UF there are 15 and 19 problems that are only solvable using the pointwise and contour encodings respectively. 
As a further point, we note that the heuristic introduced on page~\pageref{page:heuristic} is useful, without it the default pointwise approach solved 61 fewer problems.
Finally, comparing with the results in  Table~\ref{tab:results:translation}, we see that finding models for multi-sorted problems directly performs better than translating the problem to an unsorted one.

\begin{table}[t]
\caption{Experimental Results for Multi-Sorted problems.\label{tab:results:sorted}}
\centering
\begin{tabular}{l|c|cccc|ccc|c}
					 	& CVC4	&  \multicolumn{4}{c|}{Pointwise} & \multicolumn{3}{c|}{Contour}	& Without \\
					 	&		&	Default	&	Expand	&	Collapse	&	Bounds	& Default	&Collapse & Bounds	& Constraints  \\
\hline
UF: sat 			&	764 (8)	&	795	&	789	&	{\bf 901 (12)}	&	810	&886 (3)	&899 (1)& 886 (1)&154		\\
UF: unsat 			&	-	&	212	&	215	&	241	&	218	&270&261	&267&66			\\	
\end{tabular}
\end{table}



%% file: conclusion.tex

\section{Conclusions and Further Work}
\label{sec:conclusion}

We have introduced a new framework for MACE-style finite model finding for multi-sorted first-order logic. This involved two complementary SAT encodings that capture the search for a satisfying domain size assignment and techniques aimed at decreasing the size of this search space.
We have demonstrated experimentally that these techniques are effective at improving finite model finding in the unsorted setting and finding finite models for multi-sorted first-order formulas. 
Further work will consider possible extensions to uninterpreted sorts and infinite, but finitely representable, models. 
%